\documentclass[twoside,12pt,reqno]{combine}
\usepackage{amsmath,amsthm, amsfonts,amssymb}
\usepackage{color,soul}
\usepackage[dvipsnames]{xcolor}
\usepackage[all]{xy}
\usepackage{graphicx}
\usepackage{indentfirst}
\usepackage{bm}
\usepackage{mathrsfs}
\usepackage{latexsym}
\usepackage{hyperref,fancyhdr}
\usepackage{microtype}	
\usepackage{bookmark}
\usepackage[T1]{fontenc}
\usepackage[utf8]{inputenc}
\usepackage{authblk}

\usepackage{fancyhdr}
\title{Raymond Stora and Les Houches 1970}
\author{Arthur Jaffe}
\affil{Harvard University, Cambridge, MA 02138 USA\thanks{I am grateful for hospitality at the FIM of the ETH-Zurich, where this paper was written.}
}
\begin{document}
\maketitle
\begin{abstract}
The marvelous Les Houches summer school of 1970 bore the trademark of Raymond Stora.
\end{abstract}

I first met Raymond Stora when he gave a nice seminar talk on renormalization in Princeton, during the time I was starting as a graduate student.  I met him again in Paris, during the student year 1963--64 that I accompanied Arthur Wightman to the IHES.   But I only got to know Raymond during the magnificent 1970 summer school that he organized with Cecile DeWitt in Les Houches. That summer school was a ``creation'' of Raymond which  played an important role in the development of mathematical physics.  So it nicely complements Raymond's scientific inventions, and demonstrates Raymond's immense and lasting scientific vision.

In fact the 1970 Les Houches school represented a turning point for quantum field theory.  It was special, as it involved experts in several different areas, many of whom had not interacted strongly with each other.  There were not only experts in constructive field theory, experts in renormalization, and experts in mathematically-oriented statistical physics, but the participants also had among them experts in analysis, group representations, operator algebras and geometry.      Rapid progress was taking place in constructive field theory and also in statistical physics research at the time.  Bringing together leaders, students, and interested bystanders in the fields could not have been better timed. New energy emerged in both subjects over the following decade, and much of it could be traced to the synergy that developed as a consequence of interactions during Les Houches 1970.   

The meeting in Les Houches might best be described as a \textit{happening}, for it was unusual in so many different ways.  Very importantly, the school lasted practically two entire months, from 5~July to 29~August!  This time-scale meant that every participant had the opportunity to speak with, to work with, to hike with, or to dine with all the other students, postdoctoral fellows, professors, and family members who spent time there.  

Not only was the school a long one, unlike current one-week or two-week events, but every student, observer, or lecturer was expected to remain for the entire two months! This too made the occasion \textit{memorable}, in a way that I have not experienced since.  The school did not have the usual hustle-bustle of people coming and going; this relaxed schedule enabled many of the  byproducts of the school.  The extended interaction, paired with the exciting scientific developments unfolding at the time---with many of them presented at the school---gave a feeling of excitement.  In fact the conversations begun and fostered during the school helped shape the future research for numerous attendees, including persons not present at the school!  

Another unusual feature of the school was the considerable number of the ``students'' who actually were expert senior researchers.   These students were interested to learn about the new developments, for they covered a very wide intellectual spectrum, so not just by chance these experts converged on Les Houches. Interesting personalities in this category included Ron Douglas (with the young Mike Douglas in hand), John Klauder,  Jean Lascoux, Andrew L\'enard, Seymour Sherman Ray Streater, and many others.  In addition, there were three ``observers'' from the Battelle Foundation, a sponsor of the school, who sent Aldo Andreotti, Raoul Bott, and George Mackey along with their families.  Their presence added immensely to the atmosphere and covered the school with a special sheen that permeated every aspect of the event.  
The ``normal'' students included notable future leaders: Alain Connes, J\"urg Fr\"ohlich, Konrad Osterwalder, Lon Rosen, Robert Schrader, Barry Simon, Andr\'e Voros, and many others.

Les Houches 1970 had another interesting rule: the lecturers were not to leave, until they gave a copy of their notes---ready for publication.  While that created some stress for those giving lectures, it had a very good side effect: the proceedings actually appeared quickly.  The book arising from the 1970 school appeared during 1971.  With its rapid availability, the book became a standard reference. Contributions included those from Henri Epstein and Jurko Glaser on renormalization, Klaus Hepp on renormalization, Jean Ginibre on functional integration in statistical mechanics, Elliott Lieb on statistical mechanical models, Robert Griffiths on phase transitions, David Ruelle on equilibrium statistical mechanics, Oscar Lanford on functional analysis, and James Glimm and myself on quantum field models.

While the persons at the school interacted a great deal with each other, the Les Houches community remained relatively isolated.  Only one telephone (which often did not work) connected the participants by voice  with the outside.   There was no email nor internet (that idea had not yet been invented).  It is hard to think of what the world was like then, but letters remained the basic way to contact the rest of the world. 

During the school, I lived in a room in the  chalet close to the lower entrance to the school.  That chalet also housed the kitchen and the dining room on the lower level, and it had a nice view over an open field.  All the participants ate three daily meals in the restaurant.  We sat family style at long tables; conversation was easy.  We often took coffee afterwards in the coffee bar.

My room was sparse, but I recall that I had brought a scale with me, to keep track of my weight.  Near the bed was a wash basin with running water, and over it was a small mirror and light. However, I was sometimes surprised when turning on the faucet, for it seemed that I could feel an electrical current! In fact the electrical wiring sometimes gave unexpected shocks, that were transmitted elsewhere through the plumbing which served as an inadequate ground. It was even reported that one could get an occasional jolt in controlling the outdoor shower behind another chalet.  This did not seem to cause a serious problem, though it kept us alert.

But a near calamity of another sort threatened the chalet: one night it could have burned down!  The furnace in the basement provide heat all night, as the evenings were rather cold.  It had an automated stoaking mechanism, that fed coal into the iron combustion chamber.  One night that mechanism malfunctioned; it  fed more and more coal into the chamber, continually making the fire hotter and hotter, until the entire furnace glowed red.  It threatened to melt.  

Luckily the smoke that ensued woke someone in the chalet, and that person woke everyone else. We went to the basement to observe the event, and immediately knew we needed help. The secretary was at a loss, but eventually someone found the cook; he was able to telephone the fire officials who arrived and who also knew how to avert disaster.   

While that was the most dramatic incident, it was not the only one during our two month stay.  In fact we spent a good deal of time outdoors.  And one student from Holland developed a severe allergic reaction to a bee sting, which required serious medical attention.

On another occasion, the police spent most of an afternoon trying to locate some money that disappeared from the school office.  Possibly the cash had been buried in the middle of a field. At least we participants surmised this, as the police spent an afternoon walking over the field, staring at the ground in a fashion that amused us observers.  The staff would not give any explanation about what was going on.  

And as one might expect, there were several rumored new relationships budding during the time.  

Raymond Stora was central in keeping together, with his dr\^ole sense of humor, the diverse elements in these surroundings.  He seemed able to create harmony from all the diverse elements of the occasion: mathematicians, statistical physicists, quantum field theorists.  He loved to talk with the participants about the mathematical and physics topics of the school.  He also offered much advice on the beautiful mountain surroundings as well as about the lectures or the locale.  

He had the uncanny intuition that led to his bringing to Les Houches those persons who both contributed much to the school, and who also benefitted from the experience in their future work. For me, it was a lucky occasion to learn the Peierls argument in statistical physics.  A couple of years later, this turned out to be the basis for the proof of the existence of phase transitions for a $\phi^{4}$ quantum field. 

As a participant, I had never before had to lecture so intensively on complicated material.   So I was all the time working in my room, either preparing or writing lecture notes.  I recall sometimes going before breakfast to the lecture room to try out a presentation.  And by the evening I was exhausted.  For this reason, I did not interact with Raymond as much as I would have liked.  

But I admired his insight and his warmth.  I was very thrilled when, toward the end of the school, Raymond asked me to cohost  a party for all the participants.  

Two years ago I visited CERN, and stopped at Raymond's office to greet him.  He was absent on that occasion, but heard about my visit.  A month later I received a long (hand-written) letter in Cambridge, with (among other thoughts) fond recollections from our 1970 summer  in Les Houches.

\end{document}